\documentclass[11pt]{article}
\renewcommand{\baselinestretch}{1.3}
\def\be{\begin{equation}}
\def\ee{\end{equation}}
\def\ba{\begin{eqnarray}}
\def\ea{\end{eqnarray}}

\def\nn{\nonumber}
\input{epsf}
\begin{document}
\title{Probability Distribution of (Schw\"{a}mmle and Tsallis) Two-parameter Entropies \\ and the Lambert W-function}
\author{Somayeh Asgarani \thanks{email: sasgarani@ph.iut.ac.ir} \  \ and
       \ \  Behrouz Mirza \thanks{email: b.mirza@cc.iut.ac.ir } \\ \\
{ Department of  Physics, Isfahan University of Technology,
Isfahan 84156-83111,  Iran }\\}

\date{}
\vspace{2cm}
\maketitle{$\hspace{6.8cm}$\Large{Abstract}}\\
We investigate a two-parameter entropy introduced by Schw\"{a}mmle
and Tsallis and obtain its probability distribution in the
canonical ensemble. The probability distribution is given in terms
of the Lambert W-function which has been used in many branches of
physics, especially in fractal structures. Also,  extensivity of
$S_{q,q'}$ is discussed and a relationship is found to exist
between the probabilities of a composite system and its subsystems
  so that the two-parameter entropy, $S_{q,q'}$, is extensive.

\section{Introduction}
 It is  known that some physical
systems cannot be described by the Boltzmann-Gibbs(BG) statistical
mechanics. Within a long list showing power-law behaviours, we
may mention diffusion \cite{1*}, turbulence \cite{2*}, transverse
momentum distribution of hadron jets in $e^+e^-$ collisions
\cite{3*}, thermalization of heavy quarks in collisional process
\cite{4*}, astrophysics \cite{5*}, solar neutrinos \cite{6*}, and
among others \cite{7*,8*,9*,10*,11*}. Such systems typically  have
long-range interactions, long-time memory, and multifractal or
hierarchical structures. To overcome at least some of these
difficulties, Tsallis proposed a generalized entropic form
\cite{2,3,4}, namely
\be\label{11}\\
S_{q}=k\frac{1-\sum_{i=1}^{\omega}p_{i}^q}{q-1}\;,\ \ee where,
$k$ is a positive constant and $\omega$ is the total number of
microscopic states. It is clear that the q-entropy ($S_q$)
recovers the usual BG-entropy
($S_{BG}=-k\sum_{i=1}^{\omega}p_{i}\ln{p_{i}}$) in the limit
$q\rightarrow1$. By defining q-logarithm
\begin{equation}\label{12}
 \ln_qx\equiv\frac{x^{1-q}-1}{1-q}\hspace{.2cm}(\ln_1x=\ln{x})\;,
\end{equation}
the entropic form, $S_q$, can be written as
\be\label{13}\\
S_{q}=k\frac{1-\sum_{i=1}^{\omega}p_{i}^q}{q-1}=k\sum_{i=1}^{\omega}p_{i}\ln_q\frac{1}{p_{i}}\;.\
\ee Hence, in the case of equiprobability,
$p_i=\frac{1}{\omega}$, the well-known Boltzmann law is recovered
 in the limit $q\rightarrow1$. The q-entropy also satisfies the relevant properties of entropy
like expansibility, composability, Lesche-stability \cite{5+} and
concavity (for $q>0$). The inverse function of the q-logarithm is
called q-exponential \cite{7} and is given by
\begin{equation}\label{14}
\exp_qx\equiv[1+(1-q)x]^\frac{1}{1-q}\hspace{.5cm}(\exp_1x=\exp{x})\;.
\end{equation}
Recently in \cite{6}, the two-parameter logarithm $\ln_{q,q'}(x)$
and exponential $\exp_{q,q'}(x)$ were defined which recovered
q-logarithm and q-exponential, in the limit $q\rightarrow1$ or
$q'\rightarrow1$. So, the two-parameter entropy, similar to
Eq.~(\ref{11}), can be defined as
\begin{equation}\label{113}
S_{q,q'}\equiv\sum_{i=1}^{\omega}p_i\ln_{q,q'}{\frac{1}{p_i}}=
\frac{1}{1-q'}\sum_{i=1}^{\omega}p_i{\Big[}\exp{\Big(}\frac{1-q'}{1-q}(p_i^{q-1}-1){\Big)}-1{\Big]}\;.
\end{equation}
The above entropy for the whole range of the space parameter does
not fulfill all the necessary properties of a physical entropy.
Therefore, it may  not be appropriate for describing the physical
systems, but useful for  solving optimization problems. In this
paper, we will find the probability distribution $p_i$ for the
two-parameter entropy $S_{q,q'}$ (\ref{113}), when canonical
constraints are imposed on the system. As a result, it will be
shown that the probability distribution is expressed in terms of
the Lambert W-function  \cite{11,12,12-,12+}, also called omega
function, which is an analytical function of $z$  defined over
the hole complex z-plane, as the inverse function of
\begin{equation}\label{31}
z=We^W\;.
\end{equation}
Using this function, it is possible to write the series of
infinite exponents in a closed form \cite{12*}
\begin{equation}\label{33}
{{z^{z^{.^{.^{.}}}}}}=-\frac{W(-\ln{z})}{\ln{z}}\;.
\end{equation}
The above equation shows that $W$-function can be used to express
self-similarity in some fractal structures. This has been
recently shown in a number of studies. Banwell and Jayakumar
\cite{14} showed that $W$-function describes the relation between
voltage, current and resistance in a diode, Packel and Yuen
\cite{15} applied the $W$-function to a ballistic projectile in
the presence of air resistance. Other applications of the
W-function include those in statistical mechanics, quantum
chemistry, combinatorics, enzyme kinetics, vision  physiology,
engineering of thin films, hydrology, and the analysis of
algorithms \cite{16,17,18}. $S_{q}$ may become extensive in cases
where there are correlations between subsystems. Hence, a  point
to be discussed will be the
extensivity of $S_{q,q'}$ \cite{8}.\\
This paper is organized as follows. In Sec. {\bf{2}}, we will
review the Generalized two parameter entropy, $S_{q,q'}$ and its
properties. In Sec. {\bf{3}}, the probability distribution $p_i$
of the two-parameter entropy, $S_{q,q'}$, will be obtained in the
canonical formalism. In Sec. {\bf{4}}, assuming that the
two-parameter entropy be extensive, we will develop a
relationship holding between probability in the composite system
and probabilities of subsystems and in Sec. {\bf{5}}, we will
have a conclusion.

\section{Generalized two parameter entropy, $S_{q,q'}$}
In this section, we will review the procedure for finding the
two-parameter entropy, $S_{q,q'}$ \cite{6}. As we know, it is
possible to define two composition laws, the generalized q-sum and
q-product \cite{5}, defined as follows
\begin{eqnarray}
&&x\oplus_q{y}\equiv{x}+y+(1-q)xy\hspace{.2cm}\ \ \ \ \ \ \ \ \ (x\oplus_1y=x+y)\;,\label{15}\\
&&x\otimes_q{y}\equiv(x^{1-q}+y^{1-q}-1)^\frac{1}{1-q}\hspace{.2cm}\
\ \ \ \ \  \ \  (x\otimes_1y=xy)\;.\label{16}
\end{eqnarray}
Using the definition of q-logarithm (Eq.~(\ref{12})) and
q-exponential (Eq.~(\ref{14})), the above relations can be
rewritten as
\begin{eqnarray}
&&\ln_q(x y)=\ln_qx\oplus_q\ln_qy\;,\label{17}\\
&&\ln_q(x\otimes_qy)=\ln_qx+\ln_qy\;.\label{18}
\end{eqnarray}
Very recently in \cite{6}, Eqs.~(\ref{17}) and (\ref{18}) were
generalized by defining a two-parameter logarithmic function,
denoted by $\ln_{q,q'}x$, which satisfies the equation
\begin{equation}\label{19}
\ln_{q,q'}(x\otimes_qy)=\ln_{q,q'}x\oplus_{q'}\ln_{q,q'}y\;.
\end{equation}
Assuming $\ln_{q,q'}x=g(\ln_qx)=g(z)$ and using $x=y$ in
Eq.~(\ref{19}), then applying some general properties of a
logarithm function like
\begin{eqnarray}
&&\ln_{q,q'}1=0\;,\label{110}\\
&&\frac{d}{dx}\ln_{q,q'}x|_{x=1}=1\;,\label{111}
\end{eqnarray}
the two-parameter generalized logarithmic function will be given
as
\begin{equation}\label{112}
\ln_{q,q'}x=\frac{1}{1-q'}{\Big[}\exp{\Big(}\frac{1-q'}{1-q}(x^{1-q}-1){\Big)}-1{\Big]}=\ln_{q'}e^{\ln_qx}\;,
\end{equation}
and its inverse function will be defined as a two-parameter
generalized exponential, $\exp_{q,q'}x$
\begin{equation}\label{112*}
\exp_{q,q'}x={{\Big\{}1+\frac{1-q}{1-q'}\ln[1+(1-q')x]{\Big\}}}^{\frac{1}{1-q}}\;.
\end{equation}
The entropy can be constructed based on the two-parameter
generalization of the standard logarithm
\begin{equation}\label{113}
 S_{q,q'}\equiv{k}\sum_{i=1}^{\omega}p_i\ln_{q,q'}{\frac{1}{p_i}}=
 \frac{k}{1-q'}\sum_{i=1}^{\omega}p_i{\Big[}\exp{\Big(}\frac{1-q'}{1-q}(p_i^{q-1}-1){\Big)}-1{\Big]}\;,
\end{equation}
which, in the case of equiprobability
($p_i=\frac{1}{\omega}\hspace{.15cm}\forall{i}$),
$S_{q,q'}=k\ln_{q,q'}{\omega}$.\\
The above entropy is Lesche-stable and some properties such as
expansibility and concavity are satisfied if certain  restrictions
are imposed on $(q,q')$.

\section{Finding probability distribution in the canonical
ensemble} In this section, we are interested in maximizing the
entropy $S_{q,q'}$ under the constraints
\begin{eqnarray}
&&\sum_{i=1}^{\omega}p_i-1=0\;,\label{21}\\
&&\sum_{i=1}^{\omega}p_i\varepsilon_i-E=0\;.\label{22}
\end{eqnarray}
These constraints are added to the entropy with Lagrange
multipliers to construct the entropic functional
\begin{equation}\label{22}
\Phi_{q,q'}(p_i,\alpha,\beta)=\frac{S_{q,q'}}{k}+
\alpha(\sum_{i=1}^{\omega}p_i-1)+\beta\sum_{i=1}^{\omega}p_i(\varepsilon_i-E)=0\;.
\end{equation}
To reach the equilibrium state, the entropic functional
$\Phi_{q,q'}$ should be maximized, namely
\begin{equation}\label{23}
\frac{\partial\Phi_{q,q'}(p_i,\alpha,\beta)}{\partial{p_i}}=0\hspace{.15cm}\Rightarrow
\hspace{.15cm}\ln_{q,q'}\frac{1}{p_i}+p_i\frac{\partial\ln_{q,q'}(\frac{1}{p_i})}{\partial{p_i}}
+\alpha+\beta(\varepsilon_i-E)=0\;.
\end{equation}
Using the definition of the two-parameter logarithm (Eq.~
(\ref{112})) and after some calculations, we get
\begin{equation}\label{24}
\exp{{\Big(}\frac{1-q'}{1-q}({p_i}^{q-1}-1){\Big)}}
{\Big(}1-(1-q'){p_i}^{q-1}{\Big)}=
1-(1-q'){\Big(}\alpha+\beta(\varepsilon_i-E){\Big)}\;.
\end{equation}
To solve the above equation and to find $p_i(\varepsilon_i)$, one
may define
\begin{equation}\label{25}
z_i\equiv\frac{1-q'}{1-q}({p_i}^{q-1}-1)\hspace{.15cm}\Rightarrow\hspace{.15cm}
1-(1-q'){p_i}^{q-1}=(q-1)z_i+q'\;,
\end{equation}
and so,  Eq.~(\ref{24}) can be rewritten as
\begin{equation}\label{26}
(q'+(q-1)z_i)\   \exp(z_i)=\gamma_i\;,
\end{equation}
with the definition
\begin{equation}\label{27}
\gamma_i\equiv1-(1-q'){\Big(}\alpha+\beta(\varepsilon_i-E){\Big)}\;.
\end{equation}
Solving the above equation gives us
\begin{equation}\label{28}
z_i=W{\Big[}\frac{e^{\frac{q'}{q-1}}\gamma_i}{q-1}{\Big]}+\frac{q'}{1-q}\;,
\end{equation}
where, $W(z)$ is the Lambert $W$-function \cite{11,12,12-,12+}.
From Eqs.~(\ref{28}) and~(\ref{25}), the probability distribution
is given by:
\begin{equation}\label{29}
p_i=\frac{1}{Z_q}{{\Big\{}1+(1-q)W
{\Big[}\frac{e^{\frac{q'}{q-1}}\gamma_i}{q-1}{\Big]}{\Big\}}}^{\frac{1}{q-1}}\;,
\end{equation}
where,
\begin{equation}\label{29*}
Z_q=\sum_{i=1}^{\omega}{{\Big\{}1+(1-q)W
{\Big[}\frac{e^{\frac{q'}{q-1}}\gamma_i}{q-1}{\Big]}{\Big\}}}^{\frac{1}{q-1}}\;.
\end{equation}
In Eq.~(\ref{27}), $\beta$ is entered as an inverse of
 pseudo-temperature, but it may be interesting to write the probability
distribution in terms of a deformed q-exponential which is more
similar to the Boltzmann probability distribution. So, $\gamma$
in Eq. (22) can be written as
\begin{equation}\label{211}
\gamma_i={\Big(}1-\alpha(1-q'){\Big)}{\Big(}1-\frac{\beta(1-q')}{1-\alpha(1-q')}(\varepsilon_i-E){\Big)}\equiv
{\Big(}1-\alpha(1-q'){\Big)}{{\Big[}\exp_q(-\beta_q(\varepsilon_i-E)){\Big]}}^{1-q}\;,
\end{equation}
where, $\beta_q$ may be defined as the inverse of the
pseudo-temperature
\begin{equation}\label{212}
\beta_q\equiv\frac{1}{k_BT_q}\equiv\frac{\beta}{1-\alpha(1-q')}\;.
\end{equation}
 The probability distribution can be written in a better form
\begin{eqnarray}
&&p_i=\frac{1}{Z_q}{\bigg\{}\exp_q{\bigg[}W{\Big[}\frac{e^{\frac{q'}{q-1}}(1-\alpha(1-q'))}{q-1}
{{\Big(}\exp_q(-\beta_q(\varepsilon_i-E)){\Big)}}^{1-q}{\Big]}{\bigg]}{\bigg\}}^{-1}\label{213}\;.
\end{eqnarray}
\begin{figure}
\centering \epsfysize=5cm\epsfysize=5cm\epsffile{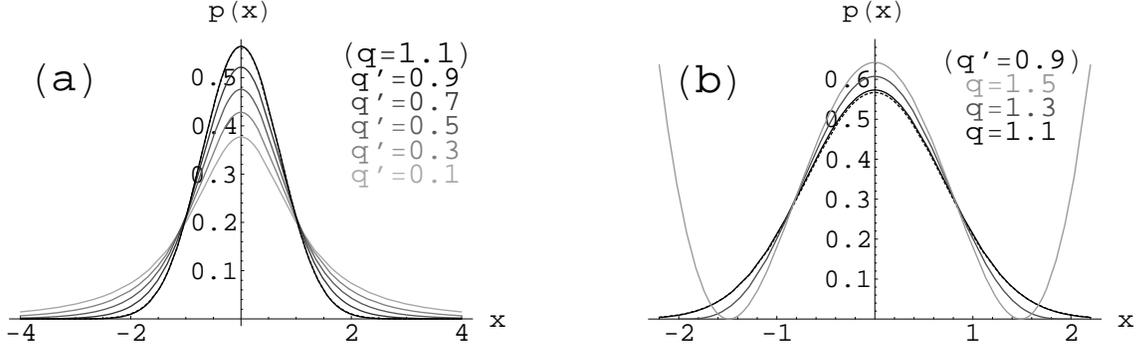}
\caption{\footnotesize Normalized probability distributions are
depicted for different values of $q$ and $q'$ in (a) and (b),
respectively. It is also shown that for $q=1.1$ and $q'=0.9$, the
plots approach to normalized Guassian. }\label{**}
\end{figure}
We can assume the energy level, $\varepsilon_i$, as a quadratic
function of the variable $x_i$. The continuous normalized
probability distribution as a function of $x$ can then be
rewritten as:
\begin{equation}\label{212}
p(x)=\frac{{{\Big\{}1+(1-q)W{\Big[}\frac{e^{\frac{q'}{q-1}}}{q-1}
{\Big(}1-(1-q')(\alpha+\beta(x^2-E)){\Big)}{\Big]}{\Big\}}^{\frac{1}{q-1}}}}
{\int_{-\infty}^{\infty}{{{\Big\{}1+(1-q)W{\Big[}\frac{e^{\frac{q'}{q-1}}}{q-1}
{\Big(}1-(1-q')(\alpha+\beta(x^2-E)){\Big)}{\Big]}{\Big\}}^{\frac{1}{q-1}}}}dx}\;.
\end{equation}
This is illustrated in Fig. 1. Comparing these plots with the
normalized Guassian distribution shows that in the limit
$q,q'\rightarrow1$, the plots approximate the Guassian function
as  expected.\\
It is also possible to repeat the procedure with the energy
constraint $\sum_{i=1}^Wp_i^q\varepsilon_i=E$. But in that case,
the equation which maximizes the entropy, is not solvable.

\section{How to interpret the entropy $S_{q,q'}$ extensive?}
Extensivity, together with concavity, Lesche stability, and
finiteness of the entropy production per time, increases the
suitability of an entropy. But, is BG-entropy the only extensive
one? In Ref.~\cite{8}, a question is raised as to whether entropy
$S_q$ is extensive or not? The answer is meaningful only if a
composition law is specified, otherwise, it is tacitly assumed
that subsystems are independent. Special correlations can be
mathematically constructed such that $S_q$
 becomes extensive for an adequate value of $q\ne1$. For example, we can consider two equal subsystems $A$ and $B$
with probabilities $p_i^A$ and $p_j^B$, respectively. The
probabilities of a composite system can be described by joint
probabilities $p_{ij}^{A+B}$. In Tables (a) and (b), two cases are
considered:
\begin{table}
\footnotesize\renewcommand{\baselinestretch}{0.4}
\begin{tabular}{c||c|c||lcr||c|c||c}
  $A\setminus{B}$ & 1 & 2 &  &  & $A\setminus{B}$ & 1 & 2 & \\ \cline{1-4} \cline{6-9}
  1 & $p_{11}^{A+B}=p^2$ & $p_{12}^{A+B}=p(1-p)$ & $p$ &  & 1 & $p_{11}^{A+B}=2p-1$ & $p_{12}^{A+B}=1-p$ & $p$\\
  \cline{1-4} \cline{6-9}
  2 & $p_{21}^{A+B}=p(1-p)$ & $p_{22}^{A+B}=(1-p)^2$ & $1-p$ &  & 2 & $p_{21}^{A+B}=1-p$ & $p_{22}^{A+B}=0$ & $1-p$ \\
  \cline{1-4} \cline{6-9}
   & $p$ & $1-p$ & 1 & &   & $p$ & $1-p$ & 1\\
  \multicolumn{9}{c}{(a)\hspace{8.1cm}(b)} \\
 \end{tabular}
 \caption{\footnotesize (a) shows the independence of two
 equal two-state subsystems A and B. The joint probabilities are
 given by the multiplication of subsystems, probabilities.
 (b) shows an example of the correlated A and B.
}\label{2}
\end{table}
(a) shows the independence of subsystems $A$ and $B$, namely
$p_{ij}^{A+B}=p_i^Ap_j^B$ and thus, the BG-entropy and the
q-entropy become extensive and non-extensive, respectively.
\begin{eqnarray}
&&\hspace{1.1cm}S_{BG}(A+B)=S_{BG}(A)+S_{BG}(B)\label{41*}\;,\\
&&S_{q}(A+B)=S_{q}(A)+S_{q}(B)+(1-q)S_{q}(A)S_{q}(B)\label{41+}\;,
\end{eqnarray}
However, (b) shows a special correlation between subsystems which
leads to non-extensivity for the BG-entropy and extensivity for
the Tsallis entropy at $q=0$.
\begin{equation}\label{41-}
S_{0}(A+B)=S_{0}(A)+S_{0}(B)\;.
\end{equation}
As can be seen in Table, (b), one of the states of the composite
system appears with the zero probability and so, the number of
effective states is $w_{eff}^{A+B}=3$, which is not equal to
$w^{A+B}=w^A\times{w^B}=4$. This simple model can be improved to
describe non-ergodic systems, where not all the states are
accessible. In the following, along the lines of  what is done in
\cite{8,9}, we will find a relation between the probabilities of
 the composite system (joint probabilities) and the probabilities
of subsystems (marginal probabilities), such that the
two-parameter entropy $S_{q,q'}$ is
extensive.\\
Now, we are interested in finding the extensivity condition for
the  two-parameter entropy. Consider $N$ subsystems
$(A_1,A_2,\ldots,A_N)$, each with the probability $p_{i_s}$ (s is
related to that system). The probabilities in the composite
system, $p_{i_1,i_2,\ldots,i_N}^{A_1+A_2+\ldots+A_N}$, should
satisfy the condition
\begin{equation}\label{41}
\sum_{i_1,i_2,\ldots,i_N}p_{i_1,i_2,\ldots,i_N}^{A_1+A_2+\ldots+A_N}=1\;.
\end{equation}
The marginal probability related to the system, $s$, is defined as
\begin{equation}\label{42}
p_{i_s}^{A_s}\equiv\sum_{i_1,i_2,\ldots{i_{s-1}}{i_{s+1}}\ldots,i_N}p_{i_1,i_2,\ldots,i_N}^{A_1+A_2+\ldots+A_N}\;.
\end{equation}
It may be interesting to find the condition which makes the
entropy $S_{q,q'}$ extensive. In other words, we want to know
 the relationship between the probability in the composite
system and probabilities of the subsystems when the entropy
$S_{q,q'}$ is extensive. Let us consider the relation
\begin{equation}\label{43}
\frac{1}{p_{i_1,i_2,\ldots,i_N}^{A_1+A_2+\ldots+A_N}}
=\exp_{q,q'}{\Big(}\sum_{s=1}^{N}\ln_{q,q'}\frac{1}{p_{i_s}^{A_s}}+\phi_{i_1,i_2,\ldots,i_N}{\Big)}\;,
\end{equation}
where, $\phi_{i_1,i_2,\ldots,i_N}$ is set to ensure
Eq.~(\ref{41}). The above equation can be rewritten in a different
form
\begin{equation}\label{44}
\frac{1}{p_{i_1,i_2,\ldots,i_N}^{A_1+A_2+\ldots+A_N}}=\frac{1}{p_{i_1}^{A_1}}
\otimes_{q,q'}\frac{1}{p_{i_2}^{A_2}}\otimes_{q,q'}\ldots\otimes_{q,q'}\frac{1}{p_{i_N}^{A_N}}
\otimes_{q,q'}\exp_{q,q'}(\phi_{i_1,i_2,\ldots,i_N})\;,
\end{equation}
with the definition of $\otimes_{q,q'}$-product \cite{6}
\begin{equation}\label{45}
x\otimes_{q,q'}y\equiv\exp_{q,q'}(\ln_{q,q'}x+\ln_{q,q'}y)\;.
\end{equation}
A nonzero function $\phi_{i_1,i_2,\ldots,i_N}$ is related to the
existence of the correlation in the system, because in the case
of independent subsystems $(q,q'\rightarrow1)$, the
$\otimes_{q,q'}$-product becomes the usual product and
$\phi_{i_1,i_2,\ldots,i_N}=0$. In Ref.~\cite{9}, the values of
$\phi_{ij}$, making the entropy $S_q$ extensive, are obtained for
two equal two-state subsystems. In the case of equiprobability,
Eq.~(\ref{44}) save for the function $\phi_{i_1,i_2,\ldots,i_N}$,
shows the generalized multiplication of the number of states of
the subsystems, which may be defined as the effective number of
states
\begin{equation}\label{45*}
w_{eff}^{A+B}=w^{A_1}\otimes_{q,q'}w^{A_2}\otimes_{q,q'}\ldots\otimes_{q,q'}w^{A_N}
\otimes_{q,q'}\exp_{q,q'}(\phi_{i_1,i_2,\ldots,i_N})\;,
\end{equation}
 The entropy of a composite system
similar to Eq.~(\ref{113}) can be defined as follows
\begin{equation}\label{46}
S_{q,q'}{\Big(}\sum_{s=1}^{N}A_s{\Big)}\equiv{k}\sum_{i_1,i_2,\ldots,i_N}
p_{i_1,i_2,\ldots,i_N}^{A_1+A_2+\ldots+A_N}\ln_{q,q'}\frac{1}{p_{i_1,i_2,\ldots,i_N}^{A_1+A_2+\ldots+A_N}}\;.
\end{equation}
Using Eq.~(\ref{43}), the entropy can be written as
\begin{eqnarray}\label{47}
&&S_{q,q'}{\Big(}\sum_{s=1}^{N}A_s{\Big)}\equiv{k}\sum_{i_1,i_2,\ldots,i_N}
p_{i_1,i_2,\ldots,i_N}^{A_1+A_2+\ldots+A_N}\ln_{q,q'}{\Big[}\exp_{q,q'}
{\Big(}\sum_{s=1}^{N}\ln_{q,q'}\frac{1}{p_{i_s}^{A_s}}+\phi_{i_1,i_2,\ldots,i_N}{\Big)}{\Big]}\nn\\
&&\hspace{2.3cm}=
k\sum_{i_1,i_2,\ldots,i_N}p_{i_1,i_2,\ldots,i_N}^{A_1+A_2+\ldots+A_N}\sum_{s=1}^{N}\ln_{q,q'}\frac{1}{p_{i_s}^{A_s}}
+k\sum_{i_1,i_2,\ldots,i_N}p_{i_1,i_2,\ldots,i_N}^{A_1+A_2+\ldots+A_N}\phi_{i_1,i_2,\ldots,i_N}\nn\\
&&\hspace{2.3cm}=
k\sum_{s=1}^{N}\sum_{i_s}p_{i_s}^{A_s}\ln_{q,q'}\frac{1}{p_{i_s}^{A_s}}
+k\sum_{i_1,i_2,\ldots,i_N}p_{i_1,i_2,\ldots,i_N}^{A_1+A_2+\ldots+A_N}\phi_{i_1,i_2,\ldots,i_N}\nn\\
&&\hspace{2.3cm}=\sum_{s=1}^{N}S_{q,q'}(A_s)+
k\sum_{i_1,i_2,\ldots,i_N}p_{i_1,i_2,\ldots,i_N}^{A_1+A_2+\ldots+A_N}\phi_{i_1,i_2,\ldots,i_N}\;,
\end{eqnarray}
where, the definition of marginal probability Eq.~(\ref{42}) is
used in the last line. Eq.~(\ref{47}) ensures extensivity of
$S_{q,q'}$ if the constraint
\begin{equation}\label{48}
\sum_{i_1,i_2,\ldots,i_N}p_{i_1,i_2,\ldots,i_N}^{A_1+A_2+\ldots+A_N}\phi_{i_1,i_2,\ldots,i_N}=0\;,
\end{equation}
is satisfied. In other words, assuming the above constraint will
be equivalent to the existence of extensivity. According to
Eqs.~(\ref{112}),~(\ref{112*}), and ~(\ref{43}), for the
probability of a composite system, we get
\begin{equation}\label{49}
p_{i_1,i_2,\ldots,i_N}^{A_1+A_2+\ldots+A_N}=
{{\Big\{}1+\frac{1-q}{1-q'}\ln{\Big[}1-N+(1-q')\phi_{i_1,i_2,\ldots,i_N}-
\sum_{s=1}^{N}\exp[\frac{1-q'}{1-q}(p_{i_s}^{q-1}-1)]{\Big]}{\Big\}}}^{\frac{1}{q-1}}\;.
\end{equation}
It is clear that in the limit $q\rightarrow1$, the proposed
relation of probabilities   is recovered \cite{8,9}
\begin{equation}\label{410}
p_{i_1,i_2,\ldots,i_N}^{A_1+A_2+\ldots+A_N}={{\Big[}1-N+(1-q')\phi_{i_1,i_2,\ldots,i_N}+
\sum_{s=1}^{N}(p_{i_s})^{q'-1}{\Big]}}^{\frac{1}{q'-1}}\;.
\end{equation}
where, in the limit $q'\rightarrow1$, the usual product of
probabilities,
$p_{i_1,i_2,\ldots,i_N}^{A_1+A_2+\ldots+A_N}=\prod_{s=1}^Np_{i_s}$,
is given, which describes the case of independent subsystems.

\vspace{.7cm}
\section{Conclusion}
In this paper, a special set of two-parameter entropies \cite{6}
were maximized in the canonical ensemble by the energy constraint
$\sum_{i=1}^{\omega}p_i\varepsilon_i=E$. We expected that the
probability distribution, $p_i(\varepsilon_i)$, can be expressed
in terms of the generalized two-parameter exponential defined in
\cite{6}. But unexpectedly, solution of  the related equation
took the form of the Lambert function which has been used in many
branches including statistical mechanics, quantum chemistry,
enzyme kinetics, and thin films, among others. The Lambert
function can also describe some fractal structures because
infinite exponents can be written in terms of the Lambert function
(Eq.~\ref{33}). This suggests that the two-parameter entropies
are probably related to the fractal structures in a phase space
which may be a subject for future study. Also, assuming extensive
$S_{q,q'}$, the probability of a composite system was given in
terms of probabilities of the subsystems.

\end{document}